\documentclass[journal]{IEEEtran}

\ifCLASSINFOpdf
\else
\usepackage[dvips]{graphicx}
\fi
\usepackage{url}

\usepackage{graphicx}

\usepackage{cite} 
\usepackage{amsmath,amsfonts,amssymb,bm}
\usepackage{algorithm,algpseudocode}
\usepackage{booktabs,multirow}
\usepackage{xcolor}
\usepackage{graphicx}
\makeatletter
\renewcommand\floatc@ruled[2]{{\color{black}{\@fs@cfont #1} #2\par}}
\renewcommand\fs@ruled{%
	\def\@fs@cfont{\bfseries\color{black}}%
	\let\@fs@capt\floatc@ruled
	\def\@fs@pre{{\color{black}\hrule height.8pt depth0pt}\kern2pt}%
	\def\@fs@post{\kern2pt{\color{black}\hrule}\relax}%
	\def\@fs@mid{\kern2pt{\color{black}\hrule}\kern2pt}%
	\let\@fs@iftopcapt\iftrue}
\makeatother
\usepackage[caption=false,font=footnotesize]{subfig}
\graphicspath{{figure/}}

\definecolor{orange}{RGB}{255,107,0}

\begin{document}
	
	\title{Deep Plug-and-Play Proximal Distance Algorithm for Robust Principal Component Analysis}
	\author{Guanhua Liu, Mingjie Shao, Wenxuan Sun, Zhi Liu, and Yanlong Zhao
	\thanks{This work was supported by the National Natural Science Foundation of China (NSFC) under Grant 62401340, 62588101,
		and the CAS Pioneer Hundred Talents Program. 
	\emph{(Corresponding author: Mingjie Shao.)}}
		\thanks{Guanhua Liu, Wenxuan Sun, and Zhi Liu are with 
			the School of Information Science and Engineering, 
			Shandong University, Qingdao 266237, China 
			(e-mail: ghLiu@mail.sdu.edu.cn; 
			wenxuansun@mail.sdu.edu.cn; liuzhi@sdu.edu.cn).}
		\thanks{Mingjie Shao and Yanlong Zhao are with the 
			State Key Laboratory of Mathematical Sciences, Academy of Mathematics and Systems Science, Chinese Academy of Sciences, Beijing 100190, China 
			(e-mail: mingjieshao@amss.ac.cn; ylzhao@amss.ac.cn).}}
	\markboth{IEEE SIGNAL PROCESSING LETTERS,~Vol.~XX, No.~X, 2026}{Liu: Deep PnP-PDA for Robust PCA}
	
	\maketitle
	
	\begin{abstract}

		Robust Principal Component Analysis (RPCA) is essential in signal processing, with applications in hyperspectral imaging, face recognition, and video background modeling. Traditional RPCA methods often suffer from insufficient utilization of data priors and struggle to capture complex image structures. 
		Conversely, advanced deep denoisers like DnCNN, FFDNet, and DRUNet excel at leveraging image data priors but lack physical model constraints and demonstrate poor robustness against structured noise in RPCA tasks.
		This paper proposes a novel Plug-and-Play (PnP) method for multi-channel RPCA, integrating both the physical generative model and cutting-edge deep priors. Extensive
		experiments on RGB and hyperspectral datasets show that PnP-PDA achieves better quantitative performance than conventional RPCA methods and state-of-the-art methods, including deep denoisers and tensor RPCA approaches, while preserving visual details.
	\end{abstract}
	

	\section{Introduction}
	\IEEEPARstart{R}{obust} principal component analysis (RPCA) is a fundamental topic in signal processing, with numerous applications in hyperspectral imaging~\cite{salut2023tensor}, robust face recognition~\cite{wright2009robust}, and video background modeling~\cite{han2023tensor}. In RPCA, an image is typically decoupled into two components: a low-rank matrix representing the background, and a sparse object, which may be a target or structured noise.
	Based on the physical model, RPCA has been utilized for principal component pursuit (PCP) for matrix low-rank recovery~\cite{candes2011robust}, stable PCP (SPCP) for noise-corrupted observations~\cite{zhou2010stable}, outlier pursuit for column-sparse corruptions~\cite{xu2012robust}, and multi-channel RPCA~\cite{lu2019tensor, goldfarb2014robust}. However, conventional RPCA often makes limited use of data priors, and the single low-rank structure cannot capture the intricate prior details of the images.
	
	
	On the other hand, recent advances in deep learning have led to powerful deep denoisers for image denoising, including DnCNN~\cite{zhang2017beyond}, FFDNet~\cite{zhang2018ffdnet}, and DRUNet~\cite{zhang2021plug}, to name a few. These denoisers are well-trained to capture the data priors of images and have demonstrated remarkable noise removal effects. 
	However, they often lack physical model information and are less robust to structured noise, which can lead to degraded performance in RPCA tasks when faced with sparse impulse noise and heavily corrupted observations.

	More recently, the trend of combining physical model methods with emerging deep priors has created a new paradigm for handling complex tasks in computational imaging. In this context, plug-and-play (PnP) and regularization by denoising (RED) have garnered increasing attention. For instance, PnP priors~\cite{venkatakrishnan2013plug, zhang2021plug} replace model-based proximal operators with pre-trained denoisers to recover fine textures. 
	RED~\cite{romano2017little} constructs an explicit regularizer from the denoiser output, providing a variational interpretation that complements the implicit PnP formulation. 
	While these approaches have been applied to various computational imaging tasks, including denoising, super-resolution, and deblurring, they have not yet been explored for the more challenging RPCA task.
	

	
	
	This letter focuses on leveraging the PnP framework for multi-channel RPCA, aiming to combine both physical models and deep priors. We propose an RPCA formulation that is implicitly regularized by data priors driven by deep denoisers.
	We apply the nonconvex $\ell_{2,p}$ norm for the structured noise removal and Schatten-$p$ norm for promoting the low-rank background for better separation. 
	For the resulting problem, we introduce a PnP method embedded in a dedicated proximal distance algorithm (PDA). PnP-PDA decouples physical-model optimization and prior regularization, and uses Nesterov extrapolation for acceleration. We conducted extensive experiments on thirteen datasets, including RGB images and hyperspectral image (HSI) datasets. The results show that PnP-PDA achieves better recovery performance than conventional RPCA methods and the compared state-of-the-art deep denoisers and tensor RPCA approaches.

	\section{Background and Problem Formulation}
	
	\subsection{Background of RPCA}
	Consider the following  robust recovery signal model
	\begin{equation}
		\boldsymbol{\mathcal{Y}} = \boldsymbol{\mathcal{X}} + \boldsymbol{\mathcal{S}} + \boldsymbol{\mathcal{N}},
		\label{eq:observation}
	\end{equation}
	where $\boldsymbol{\mathcal{Y}} \in \mathbb{R}^{M \times N \times C}$ represents the noise corrupted multi-channel observation, $\boldsymbol{\mathcal{X}} \in \mathbb{R}^{M \times N \times C}$ denotes the low-rank background component, $\boldsymbol{\mathcal{S}} \in \mathbb{R}^{M \times N \times C}$ denotes the fiber-sparse outlier noise, and $\boldsymbol{\mathcal{N}} \in \mathbb{R}^{M \times N \times C}$ denotes the additive Gaussian noise. 
	Here, $M \times N$ denotes the data spatial size and $C$ denotes the number of channels. 
	
	The objective of robust multi-channel recovery is to separate and estimate $\boldsymbol{\mathcal{X}}$ and $\boldsymbol{\mathcal{S}}$ from $\boldsymbol{\mathcal{Y}}$. This model arises in tasks such as multispectral imaging~\cite{salut2023tensor} and video surveillance~\cite{lu2019tensor}, where $\boldsymbol{\mathcal{X}}$ captures the stationary scene or spectral background and $\boldsymbol{\mathcal{S}}$ captures small objects or anomalous pixels.
	
	We first review the conventional RPCA formulation for the single-channel case, i.e., $C=1$ in \eqref{eq:observation}, which is given by
	\begin{equation}
		\min_{\boldsymbol{X},\boldsymbol{S}} \|\boldsymbol{X}\|_* + \lambda \|\boldsymbol{S}\|_1 \quad \text{s.t.} \quad \boldsymbol{Y} = \boldsymbol{X} + \boldsymbol{S},
		\label{eq:rpca}
	\end{equation}
	where $\boldsymbol{X} \in \mathbb{R}^{M \times N}$ denotes the low-rank component in the matrix case, and $\boldsymbol{S} \in \mathbb{R}^{M \times N}$ denotes the sparse outliers, $\lambda > 0$ balances the two terms, and $\|\cdot\|_*$ is the nuclear norm.
	This problem 
	can be solved by accelerated proximal gradient (APG)~\cite{ganesh2009fast}, alternating direction method of multipliers (ADMM)~\cite{lin2010augmented}, or singular value thresholding (SVT)~\cite{cai2010singular}.
	
	Beyond \eqref{eq:rpca}, the principles of RPCA have been exploited for low-rank and sparse decomposition, including SPCP for observations with mixed Gaussian noise~\cite{zhou2010stable, bouwmans2014robust}, tensor decomposition methods~\cite{lu2019tensor, cichocki2015tensor}.
	However, these conventional methods remain an issue: they often rely on physical models and typically explore less on the data priors, which may cause prior mismatch and limited recovery under severe corruptions.
	
	\subsection{Problem Formulation}
	
	To better harness both the physical model and data priors, we formulate a PnP RPCA formulation with an implicit deep prior, which reads as
	\begin{equation}
		\begin{split}
			\min_{\boldsymbol{\mathcal{X}}, \boldsymbol{\mathcal{S}}} \: & \gamma_1 \Phi(\boldsymbol{\mathcal{X}}) + \gamma_2 \|\boldsymbol{\mathcal{X}}\|_{S_p}^p + \gamma_3 \|\boldsymbol{\mathcal{X}}\|_{\text{TV}} + \lambda \|\boldsymbol{\mathcal{S}}\|_{2,p}^p \\
			\text{s.t.} 	& \quad \boldsymbol{\mathcal{Y}} = \boldsymbol{\mathcal{X}} + \boldsymbol{\mathcal{S}}.
		\end{split}
		\label{eq:ultimate_objective}
	\end{equation}
	In this formulation, $\Phi(\cdot)$ is an implicit deep prior, and $\gamma_1, \gamma_2, \gamma_3, \lambda \ge 0$ 
	balance the four regularizers. The $\ell_{2,p}$ norm 
	$\|\boldsymbol{\mathcal{S}}\|_{2,p}^p = \sum_{i,j} 
	\|\boldsymbol{\mathcal{S}}(i,j,:)\|_2^p$ for $0 < p \leq 1$ 
	enforces fiber-sparsity along the mode-$3$ fibers 
	$\boldsymbol{\mathcal{S}}(i,j,:) \in \mathbb{R}^C$. The tensor 
	Schatten-$p$ norm is 
	$\|\boldsymbol{\mathcal{X}}\|_{S_p}^p = \sum_{k=1}^3 
	\|\boldsymbol{X}_{(k)}\|_{S_p}^p$, where 
	$\|\boldsymbol{X}_{(k)}\|_{S_p}^p = \sum_m \sigma_{k,m}^p$ is the 
	matrix Schatten-$p$ norm with $\sigma_{k,m}$ being the $m$-th 
	singular value of $\boldsymbol{X}_{(k)}$. Here, 
		$\boldsymbol{X}_{(k)} = \mathrm{unfold}_k(\boldsymbol{\mathcal{X}}) 
		\in \mathbb{R}^{d_k \times (D/d_k)}$ denotes the tensor mode-$k$ 
		matricization with dimensions $(d_1,d_2,d_3)=(M,N,C)$, and 
	$D = MNC$, while the inverse operator $\mathrm{fold}_k$ restores 
	the multi-channel shape. 
	The anisotropic TV norm 
	is expressed in a channel-wise vectorized form as 
	$\|\boldsymbol{\mathcal{X}}\|_{\text{TV}} = \sum_{c=1}^C 
	(\|\boldsymbol{D}_h \boldsymbol{x}_c\|_1 + \|\boldsymbol{D}_v 
	\boldsymbol{x}_c\|_1)$. Here, $\boldsymbol{x}_c: = 
	\mathrm{vec}(\boldsymbol{\mathcal{X}}_{:,:,c}) \in \mathbb{R}^{MN}$ 
	denotes the vectorization of the $c$-th spatial channel, and 
	$\boldsymbol{D}_h, \boldsymbol{D}_v \in \mathbb{R}^{MN \times MN}$ 
	are the horizontal and vertical finite-difference matrices evaluated 
	under periodic boundary conditions.
	
	Compared with \eqref{eq:rpca}, the formulation in 
	\eqref{eq:ultimate_objective} integrates the physical model with a deep prior $\Phi(\cdot)$ through the PnP framework. Furthermore, to better promote low-rank and sparsity structures, we replace the nuclear norm and $\ell_1$ norm with the nonconvex Schatten-$p$ and 
	$\ell_{2,p}$ norms.
	However, the enhanced PnP formulation \eqref{eq:ultimate_objective} presents a critical nonconvex optimization challenge, particularly due to the coupling between the deep prior $\Phi$ and the other components.
	
	\section{Plug-and-Play Proximal Distance Algorithm}
	\subsection{PnP-PDA}
	
	We address the nonconvex coupling challenge with a novel PDA algorithm that reformulates the task as a consensus optimization problem. 
	Then, the global objective can be split into tractable subproblems that allow for efficient closed-form updates. 
	This decoupling isolates the deep prior $\Phi$, enabling the direct embedding of a deep denoiser as a modular solver.
	
	Our first step is to introduce the consensus constraint with $\boldsymbol{\mathcal{X}}_1, \boldsymbol{\mathcal{X}}_2, \boldsymbol{\mathcal{X}}_3=\boldsymbol{\mathcal{X}}$ by
	$$
	\mathcal{C} = \left\{ (\boldsymbol{\mathcal{X}}_1, \boldsymbol{\mathcal{X}}_2, \boldsymbol{\mathcal{X}}_3, \boldsymbol{\mathcal{S}}) \mid \boldsymbol{\mathcal{X}}_1 = \boldsymbol{\mathcal{X}}_2 = \boldsymbol{\mathcal{X}}_3, \; \boldsymbol{\mathcal{X}}_1 + \boldsymbol{\mathcal{S}} = \boldsymbol{\mathcal{Y}} \right\}.
	$$ 
	Then, problem \eqref{eq:ultimate_objective} is reformulated as
	\begin{equation}
		\min_{\boldsymbol{x} \in \mathcal{C}} f(\boldsymbol{x}) = \gamma_1 \Phi(\boldsymbol{\mathcal{X}}_1) + \gamma_2 \|\boldsymbol{\mathcal{X}}_2\|_{S_p}^p + \gamma_3 \|\boldsymbol{\mathcal{X}}_3\|_{\text{TV}}+ \lambda \|\boldsymbol{\mathcal{S}}\|_{2,p}^p,
	\end{equation}
	where $\boldsymbol{x} = (\boldsymbol{\mathcal{X}}_1, \boldsymbol{\mathcal{X}}_2, \boldsymbol{\mathcal{X}}_3, \boldsymbol{\mathcal{S}})$.
	Next, we transform the consensus constraint with a soft constraint, which is posed as
	\begin{equation}
		\textstyle F_{\rho}(\boldsymbol{x}) = f(\boldsymbol{x}) + \frac{\rho}{2} \mathrm{dist}^2(\boldsymbol{x}, \mathcal{C}),
	\end{equation}
	where $\mathrm{dist}(\boldsymbol{x}, \mathcal{C}) = \min_{\boldsymbol{y} \in \mathcal{C}} \|\boldsymbol{x} - \boldsymbol{y}\|_F$ is the distance function and $\rho > 0$ is a penalty parameter. 
	We majorize $\mathrm{dist}^2(\boldsymbol{x}, \mathcal{C})$ at $\boldsymbol{x}^t$ using a quadratic surrogate
	\begin{equation}
		\mathrm{dist}^2(\boldsymbol{x}, \mathcal{C}) \leq \|\boldsymbol{x} - P_{\mathcal{C}}(\boldsymbol{x}^{t})\|_F^2,
		\label{eq:mm_surrogate}
	\end{equation}
	with equality at $\boldsymbol{x} = \boldsymbol{x}^t$, where $P_{\mathcal{C}}$ denotes the projection onto $\mathcal{C}$ that takes the form $P_{\mathcal{C}}(\boldsymbol{x}^t) = (P_\mathcal{C}^{\mathcal{X}}, P_\mathcal{C}^{\mathcal{X}}, P_\mathcal{C}^{\mathcal{X}}, P_\mathcal{C}^{\mathcal{S}})$ with
	\begin{equation}
		P_\mathcal{C}^{\mathcal{X}} = \frac{1}{4}\Big(\sum_{i=1}^3 \boldsymbol{\mathcal{X}}_i^t + \boldsymbol{\mathcal{Y}} - \boldsymbol{\mathcal{S}}^t\Big), \quad P_\mathcal{C}^{\mathcal{S}} = \boldsymbol{\mathcal{Y}} - P_\mathcal{C}^{\mathcal{X}}.
		\label{eq:consensus_solution}
	\end{equation}
	Under this majorization, we propose the following PDA algorithm where at iteration $t$ it takes the form
	\begin{equation}
		\boldsymbol{x}^{t+1} = \arg\min_{\boldsymbol{x}} f(\boldsymbol{x}) + \frac{\rho}{2} \sum_{i=1}^3 \|\boldsymbol{\mathcal{X}}_i - P_\mathcal{C}^{\mathcal{X}}\|_F^2 + \frac{\rho}{2} \|\boldsymbol{\mathcal{S}} - P_\mathcal{C}^{\mathcal{S}}\|_F^2.
		\label{eq:pda_update_final}
	\end{equation}
	Since $f(\boldsymbol{x})$ is separable, problem \eqref{eq:pda_update_final} splits into four independent subproblems for $\boldsymbol{\mathcal{X}}_1, \boldsymbol{\mathcal{X}}_2, \boldsymbol{\mathcal{X}}_3$, and $\boldsymbol{\mathcal{S}}$, each admitting a form of proximal operator and can be efficiently solved. 
	\subsubsection{Proximal Mapping for $\boldsymbol{\mathcal{X}}_1$}
	The $\boldsymbol{\mathcal{X}}_1$ subproblem reads
	\begin{equation}
		\min_{\boldsymbol{\mathcal{X}}_1}\;\gamma_1 \Phi(\boldsymbol{\mathcal{X}}_1) + \frac{\rho}{2} \|\boldsymbol{\mathcal{X}}_1 - P_\mathcal{C}^{\mathcal{X}}\|_F^2.
	\end{equation}
	which amounts to derive the proximal mapping $\mathrm{prox}_{(\gamma_1/\rho)\Phi}(P_\mathcal{C}^{\mathcal{X}})$. Since $\Phi$ is an implicit data prior, its proximal subproblem can be addressed by a denoiser under the PnP framework~\cite{zhu2023denoising}, as
	\begin{equation}
		\boldsymbol{\mathcal{X}}_1^{t+1} = \operatorname{Denoiser}(P_\mathcal{C}^{\mathcal{X}}, \sigma_t),
		\label{eq:pnp_update}
	\end{equation}
	where $\sigma_t = \sqrt{\gamma_1/\rho}$ is the noise level parameter.
	\subsubsection{Proximal Mapping for $\boldsymbol{\mathcal{X}}_2$}
	The $\boldsymbol{\mathcal{X}}_2$ subproblem is
	\begin{equation}
		\min_{\boldsymbol{\mathcal{X}}_2} \; \gamma_2 \|\boldsymbol{\mathcal{X}}_2\|_{S_p}^p + \frac{\rho}{2} \|\boldsymbol{\mathcal{X}}_2 - P_\mathcal{C}^{\mathcal{X}}\|_F^2.
		\label{eq:X1_update}
	\end{equation}
	Since the $\mathcal{X}_{2}$ subproblem sums over three mode unfoldings coupled through the same tensor variable, we approximate it using a mode-wise decoupling-and-averaging strategy with three matrix subproblems for $k\in\{1,2,3\}$, written as
	\begin{equation}
		\min_{\boldsymbol{X}_{2(k)}} \; \gamma_2 \|\boldsymbol{X}_{2(k)}\|_{S_p}^p + \frac{\rho}{2} \|\boldsymbol{X}_{2(k)} - P_{\mathcal{C}(k)}^{\mathcal{X}}\|_F^2.
		\label{eq:mode_k_matrix}
	\end{equation}
	Let $P_{\mathcal{C}(k)}^{\mathcal{X}} = \boldsymbol{U}_k 
	\boldsymbol{\Sigma}_k \boldsymbol{V}_k^\top$ be the SVD of the 
	mode-$k$ matricization of $P_\mathcal{C}^{\mathcal{X}}$, where 
	$\sigma_{k,m}$ denotes its $m$-th singular value. By the unitary 
	invariance of the Schatten-$p$ and Frobenius norms, the optimal $\boldsymbol{X}_{2(k)}$ shares the singular vectors $\boldsymbol{U}_k$ and $\boldsymbol{V}_k$ of $P_{\mathcal{C}(k)}^{\mathcal{X}}$~\cite{cai2010singular}, 
and the optimization reduces 
to $d_k$ independent scalar problems for the sought singular values $\tilde{\sigma}_m \ge 0$
\begin{equation}
	\min_{\tilde{\sigma}_m \geq 0} \; \gamma_2 \tilde{\sigma}_m^p 
	+ \frac{\rho}{2}(\tilde{\sigma}_m - \sigma_{k,m})^2, 
	\quad m = 1, \ldots, d_k.
	\label{eq:scalar_sv}
\end{equation}
Since $0 < p < 1$, the term $\tilde{\sigma}_m^p = 
(\tilde{\sigma}_m^2)^{p/2}$ is non-convex. Following the reweighted 
least squares (RLS) strategy~\cite{fu2015joint}, we majorize this 
term at the current iterate by the first-order upper bound 
$\tilde{\sigma}_m^p \leq w_k^{(m)} \tilde{\sigma}_m^2 + c_k^{(m)}$, 
where $c_k^{(m)}$ is a constant and the scalar weight is
\begin{equation}
	w_{k}^{(m)} = \frac{p}{2} \left( (\sigma_m^t)^2 
	+ \epsilon \right)^{\frac{p-2}{2}},
	\label{eq:sv_weight}
\end{equation}
with $\sigma_m^t$ denoting the corresponding singular value of the 
previous iterate $\boldsymbol{X}_{2(k)}^t$ and $\epsilon > 0$ 
ensuring numerical stability. Substituting this majorizer back into 
\eqref{eq:scalar_sv} and setting the derivative to zero yields the closed-form reweighted shrinkage
\begin{equation}
	\tilde{\sigma}_m = \frac{\rho}{\rho + 2\gamma_2 w_{k}^{(m)}} 
	\,\sigma_{k,m}.
	\label{eq:sv_shrink_formula}
\end{equation}
Averaging the three mode estimates and refolding yields the multi-channel update
\begin{equation}
	\boldsymbol{\mathcal{X}}_2^{t+1} = \frac{1}{3} \sum_{k=1}^{3} \mathrm{fold}_k \left( \boldsymbol{U}_k \tilde{\boldsymbol{\Sigma}}_k \boldsymbol{V}_k^{\top} \right),
	\label{eq:sv_shrink}
\end{equation}
where $\tilde{\boldsymbol{\Sigma}}_k = \mathrm{diag}(\tilde{\sigma}_1, \ldots, \tilde{\sigma}_{d_k})$. Within each PDA iteration, the RLS performs a single reweighting step.
\subsubsection{Proximal Mapping for $\boldsymbol{\mathcal{X}}_3$}
The $\boldsymbol{\mathcal{X}}_3$-subproblem is
\begin{equation}
	\min_{\boldsymbol{\mathcal{X}}_3}\;\sum_{c=1}^{C} \left[
	\gamma_3 \|\boldsymbol{X}_{3,c}\|_{\text{TV}} 
	+ \frac{\rho}{2} \|\boldsymbol{X}_{3,c} 
	- P_{\mathcal{C},c}^{\mathcal{X}}\|_F^2
	\right],
	\label{eq:tv_subproblem}
\end{equation}
where $\boldsymbol{X}_{3,c}, P_{\mathcal{C},c}^{\mathcal{X}} \in 
\mathbb{R}^{M \times N}$ denote the $c$-th channel of 
$\boldsymbol{\mathcal{X}}_3$ and $P_\mathcal{C}^{\mathcal{X}}$, 
respectively.
Since the problem is independent across channels, we 
describe the procedure for each channel. 
Since 
$\boldsymbol{D}_h$ and $\boldsymbol{D}_v$ couple the entries of 
$\boldsymbol{x}_c$, we adopt Fast Total Variation deconvolution 
(FTVd)~\cite{wang2008new} by introducing auxiliary variables 
$\boldsymbol{g}_h = \boldsymbol{D}_h \boldsymbol{x}_c$ and 
$\boldsymbol{g}_v = \boldsymbol{D}_v \boldsymbol{x}_c$ and relaxing 
the equality constraints with a quadratic penalty of weight $\beta$,
\begin{equation}
	\begin{split}
		\min_{\boldsymbol{x}_c, \boldsymbol{g}_h, \boldsymbol{g}_v}
		\; &\gamma_3 (\|\boldsymbol{g}_h\|_1 
		+ \|\boldsymbol{g}_v\|_1) 
		+ \frac{\rho}{2} \|\boldsymbol{x}_c 
		- \mathrm{vec}(P_{\mathcal{C},c}^{\mathcal{X}})\|_2^2 \\
		&+ \frac{\beta}{2}\|\boldsymbol{g}_h 
		- \boldsymbol{D}_h \boldsymbol{x}_c\|_2^2 
		+ \frac{\beta}{2}\|\boldsymbol{g}_v 
		- \boldsymbol{D}_v \boldsymbol{x}_c\|_2^2.
	\end{split}
	\label{eq:tv_split}
\end{equation}
We choose $\beta=\rho$. For any given
$\boldsymbol{x}_c$,  the solutions for 
$\boldsymbol{g}_h$ and $\boldsymbol{g}_v$ is given by soft-thresholding
\begin{equation}
	\hat{\boldsymbol{g}}_h = \mathcal{T}_{\gamma_3/\rho}
	(\boldsymbol{D}_h \boldsymbol{x}_c), \quad 
	\hat{\boldsymbol{g}}_v = \mathcal{T}_{\gamma_3/\rho}
	(\boldsymbol{D}_v \boldsymbol{x}_c),
	\label{eq:tv_shrink}
\end{equation}
where $\mathcal{T}_{\tau}(x) = \mathrm{sgn}(x)\max(|x|-\tau,0)$. 
Substituting $\hat{\boldsymbol{g}}_h, \hat{\boldsymbol{g}}_v$ back 
into \eqref{eq:tv_split} and setting the gradient with respect to 
$\boldsymbol{x}_c$ to zero gives the normal equation 
$\boldsymbol{A}\boldsymbol{x}_c = \boldsymbol{b}$, where 
\[
\boldsymbol{A} = \boldsymbol{I} + \boldsymbol{D}_h^{\top}
\boldsymbol{D}_h + \boldsymbol{D}_v^{\top}\boldsymbol{D}_v, 
\boldsymbol{b} = \mathrm{vec}(P_{\mathcal{C},c}^{\mathcal{X}}) 
+ \boldsymbol{D}_h^{\top}\hat{\boldsymbol{g}}_h 
+ \boldsymbol{D}_v^{\top}\hat{\boldsymbol{g}}_v.\]
Under periodic 
boundary conditions, $\boldsymbol{D}_h$ and $\boldsymbol{D}_v$ are 
circulant and diagonalized by the DFT matrix 
$\boldsymbol{F}$ as 
$\boldsymbol{D}_h = \boldsymbol{F}^{H}\boldsymbol{\Lambda}_{h}
\boldsymbol{F}$ and $\boldsymbol{D}_v = \boldsymbol{F}^{H}
\boldsymbol{\Lambda}_{v}\boldsymbol{F}$. 
Then, the closed-form solution is
\begin{equation}
	\boldsymbol{x}_c = \boldsymbol{F}^{H} \boldsymbol{\Lambda}^{-1} 
	\boldsymbol{F} \boldsymbol{b},
	\label{eq:TV_update}
\end{equation}
where $\boldsymbol{\Lambda} = \boldsymbol{I} 
+ \boldsymbol{\Lambda}_{h}^{H}\boldsymbol{\Lambda}_{h} 
+ \boldsymbol{\Lambda}_{v}^{H}\boldsymbol{\Lambda}_{v}$ is a 
constant diagonal matrix independent of $\rho$ and $t$, precomputed 
once and reused across all $C$ channels.

\subsubsection{Proximal Mapping for $\boldsymbol{\mathcal{S}}$}
The $\boldsymbol{\mathcal{S}}$ subproblem reads as
\begin{equation}
	\min_{\boldsymbol{\mathcal{S}}}\; \lambda 
	\|\boldsymbol{\mathcal{S}}\|_{2,p}^p + \frac{\rho}{2} 
	\|\boldsymbol{\mathcal{S}} - P_\mathcal{C}^{\mathcal{S}}\|_F^2.
\end{equation}
Applying the same RLS scheme as in the 
$\boldsymbol{\mathcal{X}}_2$ subproblem yields the closed-form 
reweighted tubal shrinkage
\begin{equation}
	\boldsymbol{\mathcal{S}}^{t+1}(i,j,:) = \frac{\rho}{\rho 
		+ 2\lambda\, \omega_{i,j}}\, 
	P_\mathcal{C}^{\mathcal{S}}(i,j,:),
	\label{eq:s_update}
\end{equation}
where $\omega_{i,j} = \frac{p}{2} \left( 
\|\boldsymbol{\mathcal{S}}^t(i,j,:)\|_2^2 
+ \epsilon \right)^{(p-2)/2}$ is the fiber-wise weight at spatial 
location $(i,j)$.

\subsection{PDA Acceleration and Complexity Analysis}

We introduce Nesterov-type extrapolation to accelerate numerical convergence, which has been shown to effectively accelerate convergence in many optimization algorithms~\cite{keys2019proximal,nesterov2013introductory,shao2024accelerated}. At iteration $t$, the 
extrapolated point is calculated by
\begin{equation}
	\boldsymbol{\mathcal{Z}}_i^t = \boldsymbol{\mathcal{X}}_i^t 
	+ \frac{t-1}{t+2}(\boldsymbol{\mathcal{X}}_i^t 
	- \boldsymbol{\mathcal{X}}_i^{t-1}), \quad i \in \{1,2,3\},
	\label{eq:extrapolation}
\end{equation}
and similarly $\boldsymbol{\mathcal{Z}}_{\mathcal{S}}^t$ for 
$\boldsymbol{\mathcal{S}}^t$. The PDA update is then performed at 
the extrapolated point by replacing 
$\boldsymbol{\mathcal{X}}_i^t$ with $\boldsymbol{\mathcal{Z}}_i^t$ 
and $\boldsymbol{\mathcal{S}}^t$ with 
$\boldsymbol{\mathcal{Z}}_{\mathcal{S}}^t$ in the consensus 
projection \eqref{eq:consensus_solution}, while the four proximal 
subproblems \eqref{eq:pnp_update}--\eqref{eq:s_update} remain 
unchanged. We also adopt a homotopy strategy for $\rho$ by 
initializing it with a small value and increasing it via $\rho^{t+1} = \min(\kappa \rho^t, \rho_{\max})$ with $\kappa > 1$, which gradually enforces the consensus constraints. 
The complete procedure is summarized in 
Algorithm~\ref{alg:pda_pnp}. The algorithm complexity analysis is provided in the supplementary material.

\begin{algorithm}[t!]
	\caption{PnP-PDA for RPCA}
	\label{alg:pda_pnp}
	\begin{algorithmic}[1]
		\Require Observation $\boldsymbol{\mathcal{Y}}$, parameters 
		$\gamma_1, \gamma_2, \gamma_3, \lambda, p$, smoothing factor 
		$\epsilon$, scaling factor $\kappa > 1$, and $\rho_{\max}$.
		\State \textbf{Initialize:} 
		$\boldsymbol{\mathcal{X}}_i^0 = 
		\text{MedianFilter}(\boldsymbol{\mathcal{Y}})$; $\boldsymbol{\mathcal{S}}^0 = \mathbf{0}$; 
		$\rho^0 > 0$.
		\Repeat
		\State Compute extrapolated points 
		$\boldsymbol{\mathcal{Z}}_i^t, 
		\boldsymbol{\mathcal{Z}}_{\mathcal{S}}^t$ via 
		\eqref{eq:extrapolation};
		\State Compute consensus projection 
		$P_\mathcal{C}^{\mathcal{X}}, P_\mathcal{C}^{\mathcal{S}}$ 
		via \eqref{eq:consensus_solution};
		\State Update $\boldsymbol{\mathcal{X}}_1^{t+1}$ via deep 
		denoiser \eqref{eq:pnp_update};
		\State Update $\boldsymbol{\mathcal{X}}_2^{t+1}$ via RLS 
		singular value shrinkage \eqref{eq:sv_shrink};
		\State Update $\boldsymbol{\mathcal{X}}_3^{t+1}$ via TV 
		proximal operator \eqref{eq:TV_update};
		\State Update $\boldsymbol{\mathcal{S}}^{t+1}$ via RLS tubal 
		shrinkage \eqref{eq:s_update};
		\State Update penalty: $\rho^{t+1} \gets \min(\kappa \rho^t, 
		\rho_{\max})$;
		\Until{convergence}
		\State \textbf{Return} $P_\mathcal{C}^{\mathcal{X}}, \; 
		P_\mathcal{C}^{\mathcal{S}} = \boldsymbol{\mathcal{Y}} 
		- P_\mathcal{C}^{\mathcal{X}}$.
	\end{algorithmic}
\end{algorithm}

\begin{table*}[!t]
	\centering
	\renewcommand{\arraystretch}{0.68}
	\setlength{\tabcolsep}{1.35pt}
	\caption{Comparison of recovery methods on ten RGB benchmark images at $\sigma=0.10$ under $10\%$ and $20\%$ corruption.}
	\vspace{-2mm}
	\label{tab:comparison_results}
	\resizebox{\textwidth}{!}{
		\begin{tabular}{ll|cccccccccc}
			\toprule
			\begin{tabular}{c}\textbf{Method}\\\textbf{(PSNR/SSIM/LPIPS)}\end{tabular}
			& \textbf{sp}
			& \textbf{Peppers} & \textbf{Girl} & \textbf{Bell} & \textbf{House} 
			& \textbf{Plane} & \textbf{Child} & \textbf{kodim23} 
			& \textbf{kodim24} & \textbf{kodim04} & \textbf{kodim15} \\
			\midrule
			\multirow{2}{*}{IALM~\cite{lin2010augmented}}
			& 0.1 & 13.27/0.3178/0.9547 & 14.94/0.3883/0.8013 & 16.95/0.2421/1.0123 & 14.66/0.3353/0.9113 & 16.17/0.3885/0.7972 & 17.41/0.3971/0.8834 & 14.11/0.3169/1.1248 & 17.20/0.5077/0.6659 & 16.14/0.3486/0.9963 & 15.35/0.3772/0.8045 \\
			& 0.2 & 12.94/0.2770/1.0247 & 14.04/0.3221/0.9057 & 15.43/0.1898/1.1019 & 13.71/0.2784/1.0148 & 14.60/0.3243/0.9332 & 16.74/0.3401/0.9614 & 13.48/0.2616/1.1699 & 15.91/0.4171/0.8198 & 15.24/0.2869/1.1016 & 14.27/0.3227/0.8925 \\
			\midrule
			\multirow{2}{*}{SPCP~\cite{zhou2010stable}}
			& 0.1 & 12.49/0.2929/1.0147 & 14.67/0.3766/0.8182 & 16.66/0.2342/1.0262 & 14.38/0.3296/0.9140 & 16.04/0.3846/0.8061 & 16.67/0.3767/0.8993 & 13.61/0.2952/1.1334 & 16.49/0.4908/0.6871 & 15.95/0.3447/1.0100 & 15.02/0.3712/0.8155 \\
			& 0.2 & 12.22/0.2547/1.0872 & 13.82/0.3129/0.9206 & 15.22/0.1857/1.1095 & 13.50/0.2751/1.0209 & 14.47/0.3210/0.9408 & 16.05/0.3245/0.9810 & 13.06/0.2457/1.1771 & 15.37/0.4059/0.8329 & 15.05/0.2839/1.1087 & 13.97/0.3175/0.9002 \\
			\midrule
			\multirow{2}{*}{DnCNN~\cite{zhang2017beyond}}
			& 0.1 & 25.83/0.7587/0.1369 & 27.26/0.8013/0.1206 & 29.94/0.8274/0.1709 & 28.19/0.8090/0.0651 & 26.21/0.7953/0.1242 & 25.65/0.6860/0.2440 & 27.02/0.7730/0.1141 & 25.70/0.7694/0.1341 & 27.29/0.7603/0.1402 & 25.99/0.7419/0.1792 \\
			& 0.2 & 24.26/0.7060/0.1795 & 25.83/0.7539/0.1490 & 28.48/0.7817/0.2113 & 26.56/0.7633/0.1052 & 24.60/0.7320/0.1735 & 23.75/0.6002/0.3364 & 25.51/0.7182/0.1860 & 24.49/0.7100/0.1896 & 25.71/0.7059/0.2093 & 23.98/0.6706/0.2347 \\
			\midrule
			\multirow{2}{*}{FFDNet~\cite{zhang2018ffdnet}}
			& 0.1 & 23.42/0.6101/0.3313 & 23.66/0.6110/0.3404 & 23.93/0.4312/0.5661 & 23.81/0.5495/0.3103 & 23.63/0.6252/0.2765 & 23.38/0.5650/0.4761 & 23.66/0.6020/0.3518 & 22.90/0.6522/0.3123 & 23.37/0.5614/0.4574 & 23.61/0.6043/0.3600 \\
			& 0.2 & 19.93/0.4429/0.5892 & 19.78/0.4293/0.5413 & 19.70/0.2493/0.8837 & 19.76/0.3692/0.5891 & 19.73/0.4347/0.5267 & 19.75/0.3754/0.8082 & 19.88/0.4289/0.5759 & 19.59/0.4813/0.6036 & 19.65/0.3747/0.7592 & 19.87/0.4157/0.5901 \\
			\midrule
			\multirow{2}{*}{DRUNet~\cite{zhang2021plug}}
			& 0.1 & 21.48/0.5332/0.3694 & 21.45/0.5049/0.3716 & 21.52/0.3157/0.6091 & 21.61/0.4552/0.3375 & 21.58/0.5366/0.3167 & 21.30/0.4719/0.5382 & 21.36/0.5012/0.4031 & 20.87/0.5453/0.3397 & 21.27/0.4580/0.4916 & 21.63/0.5217/0.4266 \\
			& 0.2 & 18.10/0.3684/0.5966 & 17.77/0.3425/0.5523 & 17.66/0.1809/0.8780 & 17.84/0.2994/0.5787 & 18.06/0.3655/0.5360 & 17.96/0.2969/0.8167 & 17.86/0.3418/0.5814 & 17.50/0.3735/0.6331 & 17.68/0.2852/0.7559 & 18.18/0.3461/0.6105 \\
			\midrule
			\multirow{2}{*}{U-SGD~\cite{dong2023fast}}
			& 0.1 & 22.50/0.5568/0.3682 & 24.51/0.6560/0.4229 & 26.56/0.5751/0.5182 & 23.37/0.4939/0.3723 & 22.04/0.5413/0.3491 & 24.05/0.6072/0.5059 & 22.65/0.5482/0.3910 & 20.88/0.5784/0.3801 & 22.82/0.5406/0.5515 & 22.43/0.5445/0.3964 \\
			& 0.2 & 21.60/0.5041/0.4330 & 24.04/0.6205/0.4587 & 24.98/0.4961/0.5905 & 22.47/0.5083/0.4424 & 20.96/0.5169/0.4948 & 23.44/0.5701/0.5884 & 21.56/0.4902/0.4575 & 20.06/0.5090/0.5301 & 22.37/0.5055/0.5719 & 21.72/0.4980/0.4279 \\
			\midrule
			\multirow{2}{*}{t-CTV~\cite{wang2023guaranteed}}
			& 0.1 & 24.36/0.6355/0.2488 & 26.00/0.6766/0.2269 & 26.67/0.5196/0.4352 & 25.32/0.5662/0.2251 & 23.46/0.5930/0.2884 & 25.68/0.6352/0.3220 & 24.33/0.6355/0.2692 & 22.58/0.6779/0.2964 & 24.45/0.6068/0.3148 & 24.09/0.6124/0.2699 \\
			& 0.2 & 23.69/0.6019/0.2765 & 25.01/0.6393/0.2847 & 25.78/0.4754/0.4940 & 24.50/0.5304/0.2675 & 22.81/0.5579/0.3178 & 24.89/0.5974/0.3819 & 23.66/0.6032/0.3124 & 21.97/0.6403/0.3397 & 23.76/0.5691/0.3645 & 23.54/0.5831/0.2941 \\
			\midrule
			\multirow{2}{*}{Without PnP (ablation)}
			& 0.1 & 24.21/0.6155/0.2527 & 26.17/0.6959/0.2855 & 27.48/0.6481/0.4630 & 25.02/0.5659/0.2536 & 23.46/0.5871/0.2919 & 25.34/0.6232/0.3345 & 24.20/0.6324/0.3044 & 22.54/0.6601/0.2698 & 24.46/0.6104/0.3363 & 23.89/0.5906/0.2759 \\
			& 0.2 & 23.48/0.5976/0.3335 & 25.61/0.6827/0.3261 & 27.12/0.6532/0.4921 & 24.44/0.5576/0.2978 & 22.84/0.5710/0.3470 & 24.68/0.6032/0.4180 & 23.64/0.6183/0.3504 & 21.96/0.6304/0.3121 & 23.96/0.5941/0.4052 & 23.11/0.5755/0.3410 \\
			\midrule
			\multirow{2}{*}{\textbf{PnP-PDA}}
			& 0.1 & \textbf{28.09/0.8358/0.0492} & \textbf{30.36/0.8654/0.0573} & \textbf{32.75/0.9171/0.0364} & \textbf{31.02/0.8485/0.0275} & \textbf{27.01/0.8442/0.0502} & \textbf{29.00/0.8328/0.0590} & \textbf{28.29/0.8382/0.0513} & \textbf{24.96/0.7900/0.1096} & \textbf{27.99/0.7869/0.1330} & \textbf{28.30/0.8271/0.0996} \\
			& 0.2 & \textbf{27.01/0.8189/0.0543} & \textbf{29.55/0.8558/0.0800} & \textbf{31.47/0.9008/0.0461} & \textbf{29.76/0.8336/0.0307} & \textbf{25.46/0.8248/0.0732} & \textbf{27.84/0.8126/0.0897} & \textbf{26.80/0.8155/0.0712} & \textbf{23.80/0.7264/0.1068} & \textbf{26.91/0.7605/0.1299} & \textbf{26.95/0.8099/0.0975} \\
			\bottomrule
	\end{tabular}}
	\vspace{-4mm}
\end{table*}

\begin{table*}[!t]
	\centering
	\renewcommand{\arraystretch}{0.62}
	\setlength{\tabcolsep}{0.85pt}
	\setlength{\arrayrulewidth}{0.25pt}
	\caption{Comparison of HSI recovery methods under $10\%$ and $20\%$ corruption at Gaussian noise levels $\sigma=0.10$ and $\sigma=0.20$.}
	\label{tab:hsi_results}
	\resizebox{0.90\textwidth}{!}{
		\scriptsize
		\begin{tabular}{l|cc|cc|cc|cc|cc|cc}
			\toprule
			\multirow{3}{*}{\begin{tabular}{c}\textbf{Method}\\\textbf{(MPSNR/MSSIM)}\end{tabular}}
			& \multicolumn{6}{c|}{\textbf{sp=0.1}}
			& \multicolumn{6}{c}{\textbf{sp=0.2}} \\
			\cmidrule(lr){2-7}\cmidrule(lr){8-13}
			& \multicolumn{2}{c|}{\textbf{PaviaU}}
			& \multicolumn{2}{c|}{\textbf{Samson}}
			& \multicolumn{2}{c|}{\textbf{Jasper}}
			& \multicolumn{2}{c|}{\textbf{PaviaU}}
			& \multicolumn{2}{c|}{\textbf{Samson}}
			& \multicolumn{2}{c}{\textbf{Jasper}} \\
			\cmidrule(lr){2-3}\cmidrule(lr){4-5}\cmidrule(lr){6-7}
			\cmidrule(lr){8-9}\cmidrule(lr){10-11}\cmidrule(lr){12-13}
			& 0.10 & 0.20
			& 0.10 & 0.20
			& 0.10 & 0.20
			& 0.10 & 0.20
			& 0.10 & 0.20
			& 0.10 & 0.20 \\
			\midrule
			
			IALM~\cite{lin2010augmented}
			& 27.86/0.769 & 25.22/0.670
			& 30.98/0.795 & 28.34/0.707
			& 29.62/0.823 & 27.05/0.737
			& 26.18/0.692 & 24.33/0.621
			& 29.30/0.735 & 27.44/0.670
			& 27.87/0.765 & 26.06/0.696 \\
			
			SPCP~\cite{zhou2010stable}
			& 27.85/0.768 & 25.17/0.667
			& 30.92/0.794 & 28.25/0.704
			& 29.60/0.822 & 27.03/0.737
			& 26.15/0.691 & 24.27/0.618
			& 29.24/0.733 & 27.35/0.667
			& 27.84/0.765 & 26.04/0.696 \\
			
			DnCNN~\cite{zhang2017beyond}
			& 25.80/0.647 & 24.25/0.619
			& 25.48/0.568 & 23.48/0.580
			& 25.37/0.614 & 23.52/0.580
			& 23.73/0.543 & 22.81/0.551
			& 23.27/0.455 & 22.02/0.499
			& 23.38/0.522 & 22.21/0.518 \\
			
			FFDNet~\cite{zhang2018ffdnet}
			& 27.21/0.701 & 24.81/0.658
			& 28.09/0.752 & 23.94/0.659
			& 26.93/0.674 & 23.97/0.627
			& 26.02/0.695 & 23.16/0.576
			& 26.05/0.714 & 22.43/0.580
			& 25.63/0.661 & 22.48/0.559 \\
			
			DRUNet~\cite{zhang2021plug}
			& 27.58/0.730 & 24.44/0.629
			& 28.28/0.760 & 23.82/0.628
			& 27.18/0.692 & 23.64/0.595
			& 25.21/0.650 & 23.14/0.579
			& 25.60/0.681 & 22.61/0.611
			& 25.04/0.625 & 22.58/0.536 \\
			
			U-SGD~\cite{dong2023fast}
			& 30.35/0.795 & 26.85/0.627
			& 29.18/0.621 & 25.86/0.460
			& 30.20/0.774 & 26.95/0.638
			& 29.12/0.747 & 26.12/0.589
			& 27.14/0.537 & 24.72/0.415
			& 28.11/0.704 & 25.73/0.589 \\
			
			t-CTV~\cite{wang2023guaranteed}
			& 27.42/0.665 & 22.80/0.424
			& 28.80/0.621 & 24.18/0.396
			& 28.19/0.702 & 23.37/0.474
			& 26.58/0.625 & 22.23/0.395
			& 28.05/0.585 & 23.62/0.370
			& 27.33/0.665 & 22.76/0.444 \\
			
			Ablation
			& 29.69/0.822 & 22.15/0.395
			& 29.80/0.789 & 26.26/0.552
			& 30.80/0.840 & 23.85/0.500
			& 28.89/0.789 & 21.06/0.344
			& 29.28/0.778 & 24.31/0.440
			& 29.85/0.812 & 22.40/0.427 \\
			
			\textbf{PnP-PDA}
			& \textbf{31.34/0.864} & \textbf{29.38/0.812}
			& \textbf{33.86/0.876} & \textbf{29.77/0.820}
			& \textbf{31.33/0.852} & \textbf{28.08/0.757}
			& \textbf{30.46/0.830} & \textbf{28.80/0.793}
			& \textbf{32.24/0.836} & \textbf{28.83/0.795}
			& \textbf{30.09/0.804} & \textbf{27.54/0.752} \\
			\bottomrule
	\end{tabular}}
	\vspace{-2mm}
\end{table*}

\begin{figure*}[t]
	\centering
	\captionsetup[subfloat]{font=scriptsize}
	\subfloat[Original]{
		\begin{tabular}{@{}c@{}}
			\includegraphics[width=0.070\textwidth]{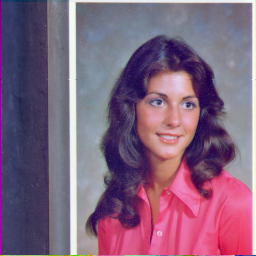} \\[-1.8mm]
			\includegraphics[width=0.070\textwidth]{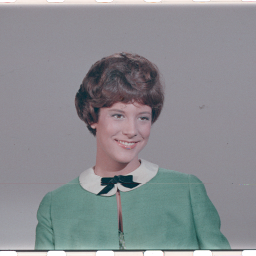}
		\end{tabular}
	}\hspace{-1.2mm}
	\subfloat[Corruption]{
		\begin{tabular}{@{}c@{}}
			\includegraphics[width=0.070\textwidth]{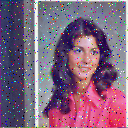} \\[-1.8mm]
			\includegraphics[width=0.070\textwidth]{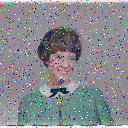}
		\end{tabular}
	}\hspace{-1.2mm}
	\subfloat[IALM]{
		\begin{tabular}{@{}c@{}}
			\includegraphics[width=0.070\textwidth]{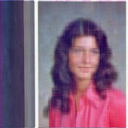} \\[-1.8mm]
			\includegraphics[width=0.070\textwidth]{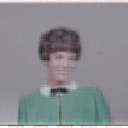}
		\end{tabular}
	}\hspace{-1.2mm}
	\subfloat[SPCP]{
		\begin{tabular}{@{}c@{}}
			\includegraphics[width=0.070\textwidth]{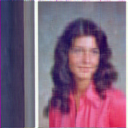} \\[-1.8mm]
			\includegraphics[width=0.070\textwidth]{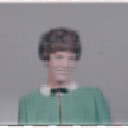}
		\end{tabular}
	}\hspace{-1.2mm}
	\subfloat[DnCNN]{
		\begin{tabular}{@{}c@{}}
			\includegraphics[width=0.070\textwidth]{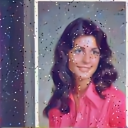} \\[-1.8mm]
			\includegraphics[width=0.070\textwidth]{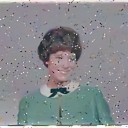}
		\end{tabular}
	}\hspace{-1.2mm}
	\subfloat[FFDNet]{
		\begin{tabular}{@{}c@{}}
			\includegraphics[width=0.070\textwidth]{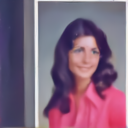} \\[-1.8mm]
			\includegraphics[width=0.070\textwidth]{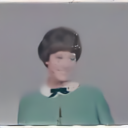}
		\end{tabular}
	}\hspace{-1.2mm}
	\subfloat[DRUNet]{
		\begin{tabular}{@{}c@{}}
			\includegraphics[width=0.070\textwidth]{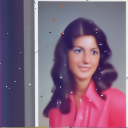} \\[-1.8mm]
			\includegraphics[width=0.070\textwidth]{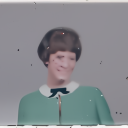}
		\end{tabular}
	}\hspace{-1.2mm}
	\subfloat[U-SGD]{
		\begin{tabular}{@{}c@{}}
			\includegraphics[width=0.070\textwidth]{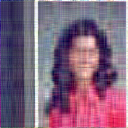} \\[-1.8mm]
			\includegraphics[width=0.070\textwidth]{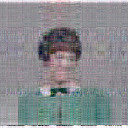}
		\end{tabular}
	}\hspace{-1.2mm}
	\subfloat[t-CTV]{
		\begin{tabular}{@{}c@{}}
			\includegraphics[width=0.070\textwidth]{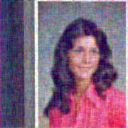} \\[-1.8mm]
			\includegraphics[width=0.070\textwidth]{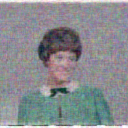}
		\end{tabular}
	}\hspace{-1.2mm}
	\subfloat[Ablation]{
		\begin{tabular}{@{}c@{}}
			\includegraphics[width=0.070\textwidth]{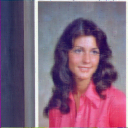} \\[-1.8mm]
			\includegraphics[width=0.070\textwidth]{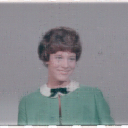}
		\end{tabular}
	}\hspace{-1.2mm}
	\subfloat[PnP-PDA]{
		\begin{tabular}{@{}c@{}}
			\includegraphics[width=0.070\textwidth]{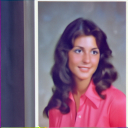} \\[-1.8mm]
			\includegraphics[width=0.070\textwidth]{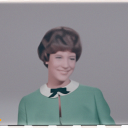}
		\end{tabular}
	}
	\caption{Visual results on ``Girl'' and ``Bell'': Original, Corruption, IALM, SPCP, DnCNN, FFDNet, DRUNet, U-SGD, t-CTV, Ablation, and PnP-PDA.}
	\vspace{-3mm}
	\label{fig:visual_comparison}
\end{figure*}


\section{Simulation Results}
\subsection{Simulation Settings and Benchmarks}
We evaluate PnP-PDA on thirteen images, including ten RGB images and three HSI datasets: PaviaU (103 bands), Samson (156 bands), and Jasper (198 bands).
All data are tested under two corruption ratios ($10\%$ and $20\%$) and Gaussian noise $(\sigma\in\{0.10,0.20\})$. 
Corrupted fibers are perturbed by i.i.d. uniform noise in $[-0.8,0.8]$. The PnP framework can plug different pre-trained denoisers for different datasets, including DRUNet~\cite{zhang2021plug} for RGB images and DPHSIR~\cite{lai2022deep} for hyperspectral images. We use PSNR, SSIM~\cite{wang2004image}, and LPIPS~\cite{zhang2018unreasonable} metrics for RGB images, and band-averaged MPSNR, MSSIM~\cite{peng2022hyperspectral} metrics for HSI. The definitions of these metrics are provided in the Information for Reproducibility.

We compare PnP-PDA against seven benchmarks and one ablation baseline. Two are conventional RPCA methods: Inexact ALM (IALM)~\cite{lin2010augmented} and SPCP~\cite{zhou2010stable}. 
Three are state-of-the-art deep denoisers: DnCNN~\cite{zhang2017beyond}, FFDNet~\cite{zhang2018ffdnet}, and DRUNet~\cite{zhang2021plug}. Two are state-of-the-art tensor RPCA methods: Unfolded ScaledGD (U-SGD)~\cite{dong2023fast} and t-CTV~\cite{wang2023guaranteed}. We added an ablation study as a baseline by examining the deep learning prior in our PnP-PDA framework.

\subsection{Performance Comparison}


Fig.~\ref{fig:visual_comparison} compares results on \textit{Girl} and \textit{Bell} at $\sigma=0.10$. Conventional RPCA methods, the ablation variant, and state-of-the-art tensor RPCA methods produce over-smoothed results, and deep denoisers leave structured artifacts, whereas PnP-PDA removes corruptions while preserving textures and edges. Extensive RGB and HSI visual results under a wider range of noise levels are provided in the supplementary material.

Table~\ref{tab:comparison_results} reports the results on ten diverse RGB images at $\sigma=0.10$. PnP-PDA achieves better performance than the state-of-the-art tensor RPCA baselines across all tests. The improvement over the ablation (without PnP) shows the contribution of the deep prior in our framework. Overall, PnP-PDA demonstrates consistently competitive quantitative performance on all ten RGB images.

Table~\ref{tab:hsi_results} reports the results on three HSI datasets with 103, 156, and 198 spectral bands. PnP-PDA achieves the highest MPSNR and MSSIM in all 12 test conditions. It consistently outperforms the conventional RPCA methods, state-of-the-art deep denoisers and tensor RPCA methods, and the ablation variant. These results support the effectiveness of PnP-PDA for HSI recovery.

\section{Conclusion}
This letter presents a PnP method for RPCA that integrates both the physical model and deep priors. Our proposed PnP-PDA algorithm transforms the problem into subproblems for evaluating proximal operators, utilizing closed-form updates and deep denoiser processing. Extensive experiments on diverse RGB and HSI datasets demonstrate the efficacy of our PnP-PDA method.



\bibliographystyle{IEEEtran}

\bibliography{references}

\end{document}